\documentclass[english,aps,graphicx,amsmath,showpacs]{revtex4}
\usepackage[T1]{fontenc}
\usepackage[latin1]{inputenc}
\usepackage{babel}
\usepackage{graphics}

\makeatletter

    \newcommand{\x}{{\mathbf {x}}} \newcommand{\W}{{\mathbf {w}}}         

\makeatother
\begin{document}

\title{Cooperating Attackers in Neural Cryptography}

\author{Lanir N. Shacham\( ^{1} \), Einat Klein\( ^{1} \), Rachel Mislovaty\( ^{1} \),
Ido Kanter\( ^{1} \) and Wolfgang Kinzel\( ^{2} \)}

\affiliation{\( ^{1} \)Minerva Center and Department of Physics, Bar-Ilan University,
Ramat-Gan, 52900 Israel,}

\affiliation{\( ^{2} \)Institut f\"ur Theoretische Physik, Universit\"at W\"urzbur,
Am Hubland 97074 W\"urzbur, Germany}

\begin{abstract}
A new and successful attack strategy in neural cryptography is
presented. The neural cryptosystem, based on synchronization of
neural networks by mutual learning, has been recently shown to be
secure under different attack strategies. The advanced attacker
presented here, named the {}``Majority-Flipping Attacker'', is the
first whose success does not decay with the parameters of the
model. This new attacker's outstanding success is due to its using
a group of attackers which cooperate throughout the
synchronization process, unlike any other attack strategy known.
An analytical description of this attack is also presented, and
fits the results of simulations.
\end{abstract}


\maketitle The use of neural networks in the field of cryptography
has recently been suggested\cite{KKK}, and has since been a source
of interest for researchers from different fields\cite{Shamir}.
The neural cryptosystem is based on the ability of two neural
networks to synchronize. The two networks undergo an online
learning procedure called \emph{mutual} \emph{learning}, in which
they learn from each other simultaneously, i.e. every network acts
both as a teacher and as a student. Every time step the networks
receive a common input vector, calculate their outputs and update
their weight vectors according to the match between their mutual
outputs \cite{MetzlerKinzelKanter}. The input/output relations are
exchanged through a public channel until their weight vectors are
identical and can be used as a secret key for encryption and
decryption of secret messages. Thus we have a public key-exchange
protocol which is not based on number theory nor involves long
numbers and irreversible functions, and is essentially different
from any other cryptographic method known before.

The question is whether this system is secure, and to what degree?
Since the data is transferred through a public channel, any
attacker who eavesdrops might manage to synchronize with the two
parties, and reveal their key. Yet the attacker is in a position
of disadvantage: while the parties perform \emph{mutual} learning
and approach one another, the attacker performs dynamic learning
and {}``chases'' them, therefore they have an advantage over him.
The system's security depends on whether they manage to exploit
this advantage so that the attacker will forever stay behind.

The synchronization is based on a competition between attractive
and repulsive stochastic forces between the parties. Attractive
forces bring them closer to each other, and Repulsive forces drive
them apart and delay the synchronization. Synchronization is
possible only if the attractive forces are stronger than the
repulsive forces, \( (A>R) \). On one hand if the attractive
forces are too strong, synchronization is relatively fast and
easy, so that an attacker eavesdropping on the line and trying to
synchronize will manage to do so easily. On the other hand if the
repulsive forces are too strong, synchronization will be hard for
the attacker, but also for the two parties. A secure system is one
which manages to balance these forces so that the net force
between the parties is positive and stronger than for the attacker
\( ((A-R)_{parties}>(A-R)_{attacker}>0) \).

The following is the model we use: The networks are Tree Parity
Machines (TPM) with \( K \) hidden units \( \sigma _{i}=\pm
1,\quad i=1,...,K \) feeding a binary output, \( \tau =\prod
_{i=1}^{K}\sigma _{i} \), as shown in Figure \ref{TPMfig}. We used
\( K=3 \). The networks consist of a discrete coupling vector \(
\W _{i}=W_{i1},...,W_{iN} \) and disjointed sets of inputs \( \x
_{i}=X_{i1},...,X_{iN} \) containing \( N \) elements each. The
input elements are random variables \( x_{ij}=\pm 1 \). Each
component of the weight vector can take certain discrete values \(
W_{ij}=\pm L,\pm (L-1),...,\pm 1,0, \) and is initiated randomly
from a flat distribution.

The local field in the \( i \)th hidden unit is defined as
\begin{equation} \label{local} h_{i}=\W _{i}\cdot\x _{i,}\end{equation}
and the output in the \( i \)th hidden unit is the sign of the
local field. The output of the tree parity machine is therefore
given by
\[\tau =\prod _{i=1}^{K}sign(h_{i})=\prod _{i=1}^{K}\sigma_{i}.\]

{
\begin{figure}
{\centering
\resizebox*{0.4\textwidth}{0.24\textheight}{\includegraphics{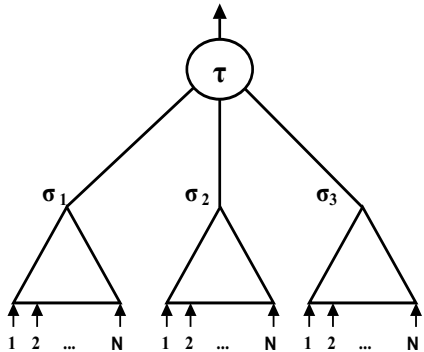}}
\par}
\vspace{-0.8cm}
\caption{\label{TPMfig}A tree parity machine with
\protect\( K=3\protect \)}
\end{figure}
\par}

During the mutual learning process, the two machines \( A \) and
\( B \), exchange their output values \( \tau ^{A/B} \). They
update their weights using the hebbian learning rule, only in case
their outputs agree, and only in hidden units which agree with the
output\begin{eqnarray}
\W _{i}^{A}(t+1)=\W _{i}^{A}(t) + \x _{i}\tau ^{A}\theta (\tau ^{A}\sigma _{i}^{A})\theta (\tau ^{A}\tau ^{B}) &  & \label{Wdisc} \\
\W_{i}^{B}(t+1)=\W_{i}^{B}(t)+\x_{i}\tau^{B}\theta(\tau^{B}\sigma_{i}^{B})\theta(\tau^{A}\tau^{B}).&\nonumber
\end{eqnarray}
This leads them to a parallel state in which \( W^{A}=W^{B} \).
The attacker, \( C \), tries to learn the weight vector of one of
the two machines, say \( A \), yet unlike the simple
teacher-student scenario\cite{EnvB,Kinzel}, the teacher's weights
in this case are time-dependant, therefore the attacker must use
some attack strategy in order to follow the teacher's steps.

The following are possible attack strategies, which were suggested
by Shamir et al.\cite{Shamir}: \emph{The Genetic Attack}, in which
a large population of attackers is trained, and every new time
step each attacker is multiplied to cover the \( 2^{K-1} \)
possible internal representations of \( \{\sigma _{i}\}\) for the
current output \( \tau \). As dynamics proceeds successful
attackers stay while the unsuccessful are removed. \emph{The
Probabilistic Attack}, in which the attacker tries to follow the
probability of every weight element by calculating the
distribution of the local field of every input and using the
output, which is publicly known. \emph{The Naive Attacker,} in
which the attacker imitates one of the parties. The most
successful attacker suggested so far is the \emph{Flipping Attack}
(Geometric attack), in which the attacker imitates one of the
parties, but in steps in which his output disagrees with the
imitated party's output, he negates (\char`\"{}flips\char`\"{})
the sign of one of his hidden units. The unit most likely to be
wrong is the one with the minimal absolute value of the local
field, therefore that is the unit which is flipped.

\begin{figure}
{\centering
\resizebox*{0.45\textwidth}{0.27\textheight}{\includegraphics{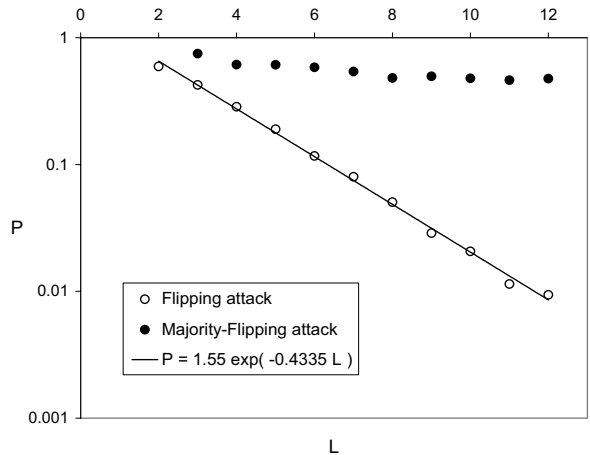}}
\par}
\caption{\label{MajAttackSuccess}The attacker's
success probability \(P\) as a function of L, for the flipping
attack and the majority-flipping attack, with N=1000, M=100,
averaged over 1000 samples. To avoid fluctuations, we define the
attacker successful if he found out 98\% of the weights}
\end{figure}

While the synchronization time increases with \( L^{2}
\)\cite{Rachel and Yaacov}, the probability of finding a
successful flipping-attacker decreases exponentially with \( L \),
\[P\propto e^{-yL}\]
as seen in Figure \ref{MajAttackSuccess}. Therefore, for large \(
L \) values the system is secure\cite{Rachel and Yaacov}. This can
be supported also by the fact that close to synchronization the
probability for a repulsive step in the mutual learning between \(
A \) and \( B \) scales like \( \left( \epsilon \right) ^{2} \),
while in the dynamic learning between the naive attacker \( C \)
and \( A \) it scales like \( \epsilon  \), where we define \(
\epsilon =Prob\left( \sigma _{i}^{C}\neq \sigma _{i}^{A}\right)
\) \cite{michal and Einat}.

The attackers mentioned above try to imitate the parties, each
using different heuristics. They use an ensemble of independent
attackers. These attackers all develop an overlap with the parties
during the synchronization process and also an overlap between
themselves, {\it yet each attacker evolves independently, and is
not influenced by the state of the other attackers}.

It has been shown that among a group of Ising vector students which
perform learning, and have an overlap R with the teacher, the best student
is the center of mass vector (which was shown to be an Ising vector
as well) which has an overlap \( R_{cm}\propto \sqrt{R} \) , for
\( R\in [0:1] \)\cite{M. Copelli}. Therefore letting the attackers
cooperate throughout the process may be to their advantage.

The new {}``Majority Flipping Attacker'' presents a general
strategy which can be applied to some of the heuristic attackers
mentioned, and improve their results, and it uses the attackers as
a {\it cooperating group rather than as individuals}, an approach
which hasn't been done before. The majority strategy is the
following: we start with a group of M random attackers. Instead of
letting them work independently and hope for one to be successful,
we let them cooperate - when updating the weights, instead of each
machine being updated according to its own result, all are updated
according to the majority's result. This {}``team-work'' approach
improves the attacker's performance. Naturally, we chose to apply
it to the most successful attacker, the {}``Flipping Attacker'',
thus creating the {}``Majority-Flipping Attacker''.

The main result of this paper is the improvement of the success
rate of the flipping attacker when using the majority scheme: The
regular flipping attacker, although relatively successful, is
weakened by increasing L, and the probability for a successful
attacker, \(P\), drops exponentially with L\cite{Rachel and
Yaacov}. When using the majority scheme, this probability seems to
approach a constant value \( \sim 0.5 \) independent of
L\cite{AntiHebb}.

When applying the majority strategy to the flipping attack, we create
M flipping attackers. In the beginning of the process, during a certain
time, the regular flipping attack is performed; those among the M
machines that disagree with party A, have one of their hidden units'
sign negated, and then their weights' vectors updated according to
their new internal representations.

After a certain time, we start to perform the majority procedure:
In every odd time step we perform the regular flipping attack, and
in every even time step we perform a majority-flipping procedure, which
consists of the following two steps:

1) All attackers who disagree with party A flip one of their
hidden units, according to the regular flipping attack procedure.

2) Now all the M attackers have the same output, but different
internal representations of \( \{\sigma _{i}\} \). We check which
of the four possible internal representations appears the most.
Then, instead of updating every attacker according to its own
internal representation, all are updated according to the same
internal representation - the majority's representation. It is as
if we let the machines \char`\"{}vote\char`\"{}, and all must use
the internal representation that was \char`\"{}elected\char`\"{}.

When the attackers perform the majority step, they all perform the
same step, therefore an overlap is developed between them. The
larger the overlap between them - the less effective they are,
because effectively there are less attackers. In the limit when
all the attackers are identical there is effectively only one
attacker. There is no way to avoid this similarity between them.
We rather prevent it from developing too quickly, and we do so by
performing the majority step only on even time steps, and not from
the beginning of the process but after a waiting time of about
1/3 of the entire synchronization time.

\begin{figure}
{\centering
\resizebox*{0.4\textwidth}{0.24\textheight}{\includegraphics{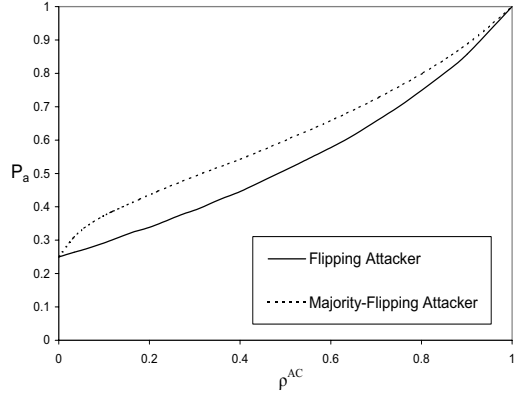}}
\par}
\caption{\label{Pa}The probability of
attacker C to have a correct internal representation as a function
of the average overlap between the attackers and one of the
parties, for flipping and majority-flipping attacks, measured in
simulations with N=1000, M=300, averaged over \( 10^{5} \)
samples.}
\end{figure}

The result of using this scheme is shown in Figure
\ref{MajAttackSuccess}. When comparing the success of the flipping
attacker with and without the majority strategy, we see that for
the latter the success probability drops exponentially with L,
while for the former it remains around 0.5 even when L is
increased. Similar results of the majority-flipping attack success
where obtained in the case of the chaotic neural network
model\cite{Chaos}.

Why is the majority-flipping attack so successful? Every update of
the weights can either bring every attacker closer to party A (an
\char`\"{}attractive step\char`\"{}) or further away (a
\char`\"{}repulsive step\char`\"{}). A repulsive step between the
attacker and A occurs when there is a difference in their internal
representations (in steps where A and B perform an update). A good
attack strategy is one that manages to reduce the probability for
a repulsive step, and the majority-flipping attacker does this by
using the majority vote. Once an overlap is developed between an
attacker and machine A, the probability for a correct (attractive)
internal representation \( P_{a} \) is larger than the probability
for a repulsive one. For a group of \( M\gg 1 \) uncorrelated
attackers, which all have an overlap \( \rho_{AC} \) with A, the
probability that their majority is correct is 1. However, if the
attackers are correlated, which is the case here, \( P_{a}<1 \),
yet it is larger than \( P_{a} \) of just one flipping attacker,
as can be seen in Figure \ref{Pa} (in our simulations we obtained
similar results for all M>50). The majority's advantage over a
random choice is the essence of this attack, as shown also in the
Bayes optimal classification algorithm vs. Gibbs learning
algorithm, where choosing the majority proves to be better then a
random choice\cite{M. Copelli}.

\begin{figure}{\centering
\resizebox*{0.4\textwidth}{0.24\textheight}{\includegraphics{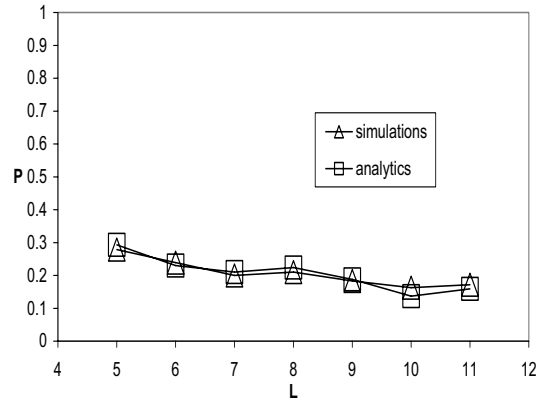}}
\par}
\caption{\label{AnalyticaGraph} The probability
for one of the M attackers to be successful as a function of L,
obtained from the analytical calculations and simulations with
N=1000, M=100. Here we define synchronization when the average
mutual overlap of the 3 hidden units reaches 0.99. Results were
averaged over 1000 samples.}
\end{figure}

The semi-analytical description of this process confirms these
results and gives us further insight to the majority attacker's
success. In the semi-analytical description we describe the system
using \( (2L+1)\times (2L+1) \) order parameters, and we manage to
simulate the system in the thermodynamic limit. We represent the
state of the TPMs using a matrix \( \mathbf{F} \) of size \(
(2L+1)\times (2L+1) \), as described in \cite{michal and Einat}.
The elements of \( \mathbf{F} \) are \( f_{qr} \), where \(
q,r=-L,...-1,0,1,...L \). The element \( f_{qr} \) represents the
fraction of components in a weight vector in which the \( A \)'s
components are equal to \( q \) and the matching components of \(
B \) are equal to \( r \). Hence, the overlap between the two
units and the norm of party A, for instance, are given by:
\begin{equation} \label{local}
R=\sum_{q,r=-L}^{L}qrf_{qr}\;\;\;Q_{A}=\sum_{q=-L}^{L}q^{2}f_{qr}
\end{equation}
and the overlap \( \rho_{AB} =R_{AB}/\sqrt{Q_{A}Q_{B}} \). There
are three matrices representing the mutual overlap between a pair
of hidden units among A, B and C (we omitted the hidden unit's
index for the sake of simplicity). We do not create M attackers
but rather one, who represents one of the M attackers in the
simulations.

This is the procedure every time step:

1)We randomly choose \( K \) local fields for the \( K \) hidden
units of machine A, from a Gaussian distribution with the mean 0
and the standard deviation \( \sqrt{Q_{A} } \).

2)We then randomly choose \( K \) local fields for the \( K \)
hidden units of machine B, from a Gaussian distribution with the
mean \( R_{AB}h_{A}/Q_{A} \) and the standard deviation \(
\sqrt{Q_{B} -R^{2}_{AB}/Q_{A}} \) (taking into account B's overlap
with A).

3)If the outputs of A and B disagree, they are not updated and we
continue to the next time step. If they agree, we update the
matrices representing A and B, and then update the attacker as
described in the next step.

4)We set the internal representation of the attacker. For K=3
there are 8 possible internal representations. We calculate their
probabilities \( P_{1},...P_{8} \), according to the attacker's
overlap with A and B and the local fields of A and B. For example
the internal representation +++ has the probability:

\[ P(+++)=\prod_{m=1}^{3} P(h^{C}_{m}>0|h^{A}_{m},h^{B}_{m},\{ R,Q \})\]

For simplicity we assume that there is no significant difference
between the attacker's overlap with A and its overlap with B and
therefore we use only one of them so that \[ P\left(
h^{C}_{i}>0\left| h^{A}_{i},\left\{ R,Q\right\} \right. \right)
=H\left( \frac{-R_{AC}h_{A}}{\sqrt{Q_{A}^{2}Q_{C}
-R_{AC}^{2}Q_{A}}}\right) \]

\noindent where \(H(x)=\int ^{\infty
}_{x}e^{-t^{2}/2}dt/\sqrt{2\pi}\).

Next we simulate the flipping, when the 8 possible states are
reduced to 4: either states 1-4 (states with positive output) flip
to 5-8 (states with negative output) or vice versa, depending on
A's output. We calculate the probabilities of the states'
flipping, for example the probability that state +++ flipped to
state -++ is: \( P\left( +++\right) \cdot P\left(
h^{C}_{1}<h^{C}_{2},h^{C}_{1}<h^{C}_{3}\right)  \) where\[
\begin{array}{c}
P\left( h^{C}_{1}<h^{C}_{2},h^{C}_{1}<h^{C}_{3}\right) =\int ^{\infty }_{0}P\left( h^{C}_{1}|h^{A}_{1},h^{B}_{1},\right) dh^{C}_{1}\cdot \\
\: \: \int ^{\infty }_{h^{C}_{1}}P\left( h^{C}_{2}|h^{A}_{2},h^{B}_{2},\right) dh^{C}_{2}\cdot \int ^{\infty }_{h^{C}_{1}}P\left( h^{C}_{3}|h^{A}_{3},h^{B}_{3},\right) dh^{C}_{3}
\end{array}\]

We now remain with probabilities for four possible internal
representations. In the case of a regular flipping step, we
randomly choose one of these four states according to their
probabilities, but in the case of a majority step, the probability
of choosing the correct internal presentation is higher. We do not
calculate it, but rather measure it in the simulations, and use
the measured probability (presented by dashed line in Figure
\ref{Pa}) in the analytical procedure. Figure \ref{AnalyticaGraph}
shows the success probability of one of the M attackers, as
function of L. It shows a fairly good agreement between the
analytical and the simulation results (see \cite{note1}).

To conclude, an important step in the field of neural cryptography
has been made, presenting a new attacking approach, under which
the TPM cryptosystem is insecure. The question is, can we create a
more sophisticated system that will be secure under the majority
attack? A secure system will be one for which the probability for
a correct step of the majority flipping attacker will be near the
flipping attacker's curve in Figure \ref{Pa}, yet the
synchronization time of the parties will still remain polynomial
with L. There can be many ideas for such kind of a system, for
example a system in which \( K>3 \), so that repulsive forces are
stronger. Yet keeping the synchronization time polynomial with L
is not easy when repulsive forces are too strong, so these models
are still under consideration, and the challenge is still
standing.


\begin{thebibliography}{10}
\bibitem{AntiHebb}The anti-hebbian learning case \cite{KKK} presents
strong finite size effects and lower \(P_{majority}\) even for
N=10000. However results seem to converge to the hebbian case in
the thermodynamic limit.
\bibitem{note1}The difference between the majority-flipping attack's success presented in Figure \ref{MajAttackSuccess}
 \( (\sim 0.5 \)) and in Figure \ref{AnalyticaGraph}  \( (\sim 0.25 \)) is due
to a different halting condition. In Figure \ref{MajAttackSuccess}
 synchronization is defined for the parties as 100\% identical weights, and 98\% identical weights for the attacker.
 In Figure \ref{AnalyticaGraph}, in order to be able to compare between the analytical and the simulations results,
 we defined synchronization as overlap=0.99
\end{thebibliography}
\end{document}